\documentclass[journal,article,submit,moreauthors,pdftex,10pt,a4paper,universe]{mdpi} 
\usepackage{physics}
\usepackage{graphicx}
\usepackage{amsmath,amssymb,amsthm,dsfont,bm}
\usepackage{color}
\usepackage{soul}
\usepackage{cancel}
\usepackage{braket}
\firstpage{1} 
\articlenumber{}
\doinum{}
\pubvolume{}
\pubyear{}
\copyrightyear{}
\externaleditor{}
\history{}

 \theoremstyle{mdpi}
 \newcounter{thm}
 \newcounter{ex}
 \newcounter{re}
\Title{Light propagation through nanophotonics wormholes}

\Author{Carlos Sab\'in}

\address{%
\quad Instituto de F\'isica Fundamental, CSIC,
Serrano, 113-bis,
28006 Madrid, Spain; csl@iff.csic.es}

\abstract{We consider the propagation of light along a 3D nanophotonic structure with the spatial shape of a spacetime containing a traversable wormhole. We show that waves experience significant changes of phase and group velocities when propagating along this curved space. This experiment can be realized with state-of-the-art nanophotonics technology.
}

\keyword{metamaterials, analogue gravity, wormholes}

\begin{document}



\setcounter{section}{-1} 
\section{Introduction}
 Optical metamaterials with non-trivial spatial dependences in the refractive index can be designed with nanophotonic structures \cite{shoresh,nikolaev,malka}. An important application is the design of nanophotonic devices where the propagation of electromagnetic fields in the visible regime mimics propagation on curved spacetime \cite{leonhardt}, thus enabling the study of general relativity properties in the laboratory \cite{experiments, experiments2}. A recent development \cite{rivkanatphot} has been the experimental design of a 3D nanophotonic material emulating Flamm's paraboloid, which  is a 2D spatial cut of the Schwarzschild spacetime metric embedded as a revolution surface in a 3D space. While, of course, this metric is perhaps the most paradigmatic example of curved spacetime due to its applications in the analysis of black holes, embedding techniques are a standard tool in general relativity and can be used to describe other spacetimes of interest \cite{misnerthornewheeler}.

An interesting example of curved spacetime is Ellis metric \cite{ellis}, which describes a spacetime containing a traversable wormhole. Unlike Einstein-Rosen bridges appearing for some observers in Schwarzschild metric -which are non-traversable- the topological shortcut of the Ellis metric can be  physically travelled through \cite{morristhorne}. So far, we don't have any experimental evidence of their existence and observational-based bounds on their abundance can be inferred \cite{search}.  The theoretical implications of the existence of traversable wormholes would entail a challenge to our understanding of causality \cite{morristhorne,morristhorne2,hawking,deutsch}. Indeed, Hawking posed a ``chronology protection conjecture'' \cite{hawking} within the semiclassical formalism of quantum field theory in curved spacetime, according to which quantum effects would prevent the appearance of closed timelike curves in spacetimes similar to Ellis, thus preventing the creation of a ``time machine". This conjecture cannot be totally confirmed or refuted in the absence of a full quantum theory of gravity. Classically, traversable wormholes require exotic energy sources violating the weak energy condition \cite{morristhorne2} and there are also tight constraints in the form of ``quantum inequalities" \cite{quantuminequalities}. However, as unlikely as their existence might look like it cannot be completely ruled out on theoretical grounds. On the other hand, it has been shown that typical phenomena commonly attributed to black holes can be perfectly mimicked by Ellis wormholes and other exotic objects. In particular, if wormholes exist even the actual identity of the objects in the center of the galaxies might be questioned \cite{combi} as well as the source of the observed gravitational waves \cite{gravastar,konoplya}. Moreover, the existence of closed timelike curves would have a significant impact in the theory of classical and quantum computing \cite{openctcs}, not to mention that wormholes play a key role in Maldacena's ``EPR-ER'' conjecture \cite{maldacena}. For all these reasons, among others, there is a growing interest both in the theoretical description of wormholes \cite{geodesics, taylor,mapping} and in their detection by classical means such as gravitational lensing \cite{lensing, lensing2}, and others \cite{shadows}, including quantum metrology techniques \cite{qdwh}. Alternatively, classical simulators \cite{rousseax,rousseax2,magneticwormhole} and quantum simulators of 1+1 dimensional reductions of Ellis and related wormhole geometries have been proposed, for instance in superconducting circuits \cite{sabinwh}, Bose-Einstein condensates \cite{sabinmateos} and trapped ions \cite{mapping}.

In this work, we propose a classical simulation of the spatial structure of a wormhole by analyzing the propagation of electromagnetic waves in the optical regime along a 3D nanophotonic structure with the shape of the embedding diagram of a traversable wormhole Ellis spacetime \cite{morristhorne2}. As in the Schwarzschild case, the embedding diagram is a 2D spacelike cut of the full spacetime embedded as a surface of revolution in 3D space. The construction of this structure would require similar technology as the Flamm's paraboloid, which has been recently build up in the laboratory \cite{rivkanatphot}. By considering realistic experimental parameters, we show that the phase and group velocities of waves propagating along the wormhole structure are significantly modified.

\section{Traversable wormhole spacetimes}
  Let us start by introducing the family of spacetime metrics considered in this work. A traversable wormhole spacetime can be characterised by \cite{morristhorne2}:
\begin{equation}
ds^2=-c^2\,e^{2\Phi(r)}dt^2+\frac{1}{1-\frac{b(r)}{r}}\,dr^2 +r^2(d\theta^2+\sin^2\theta d\phi^2), \label{eq:metric}
\end{equation}
where the {\it redshift function} $\Phi(r)$ and the {\it shape function} $b(r)$ are functions of the radius $r$ only. There is a value $b_0$ of $r$ at which $b\, (r=b_0)=r=b_0$, which determines the position of the wormhole's throat. Then, the {\it proper} radial distance to the throat is defined by \cite{morristhorne2} $l=\pm\int_{b_0}^r\,dr'(1-b(r')/r')^{-1/2}$, defining two different ``Universes'' or regions within the same Universe for $l >0$ (as $r$ goes from $\infty$ to $b_0$) and $l<0$ (as the non-monotonic $r$ goes back from $b_0$ to $\infty$). Thus, as $r\rightarrow\infty$ we have two asymptotically flat spacetime regions $l\rightarrow\pm\infty$ connected by the wormhole throat at $l=0$ ($r=b_0$). 

In this work, we will consider for simplicity that $\Phi(r)=0$ (massless wormhole). The properties of the wormhole will depend on the form of the shape function $b(r)$. In particular, as shown in \cite{morristhorne2} the parameters of this function can be adjusted in order to make traversability possible and convenient. We will consider some particular shape function later.


\section{Embedding wormhole surfaces}
We now use standard embedding techniques to obtain the shape of the embedding of a wormhole slice. Let us consider a particular instant of time and $\theta=\pi/2$ -note the spherical symmetry of the spacetime. Then we have:

\begin{equation}
ds^2=-\frac{1}{1-\frac{b(r)}{r}}\,dr^2 +r^2 d\phi^2, \label{eq:embed}
\end{equation}
This 2D spatial cut can be embedded in a 3D space as a revolution surface, given by a certain function $Z(r)$. Thus, in cylindrical coordinates the metric of the embedding would be:
\begin{equation}
ds^2=(1+(\partial_r Z(r))^2) dr^2+r^2d\phi^2.
\label{eq:embedmetric}
\end{equation}
Comparting Eqs. (\ref{eq:embed}) and (\ref{eq:embedmetric}), we get:
\begin{equation}
1+(\partial_r Z(r))^2=\frac{1}{1-\frac{b(r)}{r}},
\label{eq:embedwhgen}
\end{equation} 
which leads to:
\begin{equation}
\partial_r Z=\sqrt{\frac{b(r)}{r-b(r)}}.\label{eq:embedwhgen2}
\end{equation}
Now, we can particularize to the paradigmatic Ellis wormhole \cite{morristhorne2}:
\begin{equation}
b(r)=\frac{b_0^2}{r}\label{eq:elliswhb}.
\end{equation}
Plugging Eq. (\ref{eq:elliswhb}) into Eq. (\ref{eq:embedwhgen2}) and integrating, we get:
\begin{equation}
Z(r)=b_0\int^r_{b_0}\frac{1}{\sqrt{r^2-b_0^2}}\,dr=b_0\log{\left(\frac{r}{b_0}+\sqrt{\frac{r^2}{b_0^2}-1}\right)}=b_0\,\operatorname{arccosh}{\frac{r}{b_0}},
\label{eq:zder}
\end{equation}
where in the last step we have used the trigonometric identity $\log{\left(x+\sqrt{x^2-1}\right)}=\,\operatorname{arccosh}{x}$. Then, finally:
\begin{equation}
\frac{r}{b_0}=\cosh\left(\frac{Z}{b_0}\right).
\label{eq:rdez}
\end{equation}
\section{Surfaces of revolution in the laboratory}

As shown in \cite{rivkanatphot}, there are always coordinates $(z,x)$ such that the metric of the revolution surface can be written as:
\begin{equation}
ds^2= dz^2+\gamma(z) dx^2,
\label{eq:revsufmet}
\end{equation}
where $x=R\phi$, $R$ being the radius of the surface at $z=0$ -in our case $R=b_0$- and 
\begin{equation}
\gamma(z)=\left(\frac{\alpha(z)}{R}\right)^2,
\label{eq:gammaz}
\end{equation}
where $\gamma(z)$ is given by the parametrization of the surface:
\begin{equation}
(\alpha(z)\cos\phi,\alpha(z)\sin\phi,\beta(z)).
\label{eq:parametriz}
\end{equation}
The idea is to identify $z$ with the axis of symmetry of the surface. Therefore, from Eq. (\ref{eq:rdez}):
\begin{equation}
\left(b_0\cosh\left(\frac{z}{b_0}\right)\cos\phi,b_0\cosh\left(\frac{z}{b_0}\right)\sin\phi,z\right).
\label{eq:parametrizwh}
\end{equation}
Comparing Eq.(\ref{eq:parametriz}) and Eq. (\ref{eq:parametrizwh}) we get:
\begin{equation}
\alpha(z)=b_0\cosh{\left(\frac{z}{b_0}\right)},
\label{eq:alpha}
\end{equation}
and finally, using Eq. (\ref{eq:gammaz}):
\begin{equation}\label{eq:gammazdef}
\gamma(z)=\cosh^2{\left(\frac{z}{b_0}\right)}.
\end{equation}

\begin{figure}[h!]\centering
\includegraphics[width=0.7\textwidth]{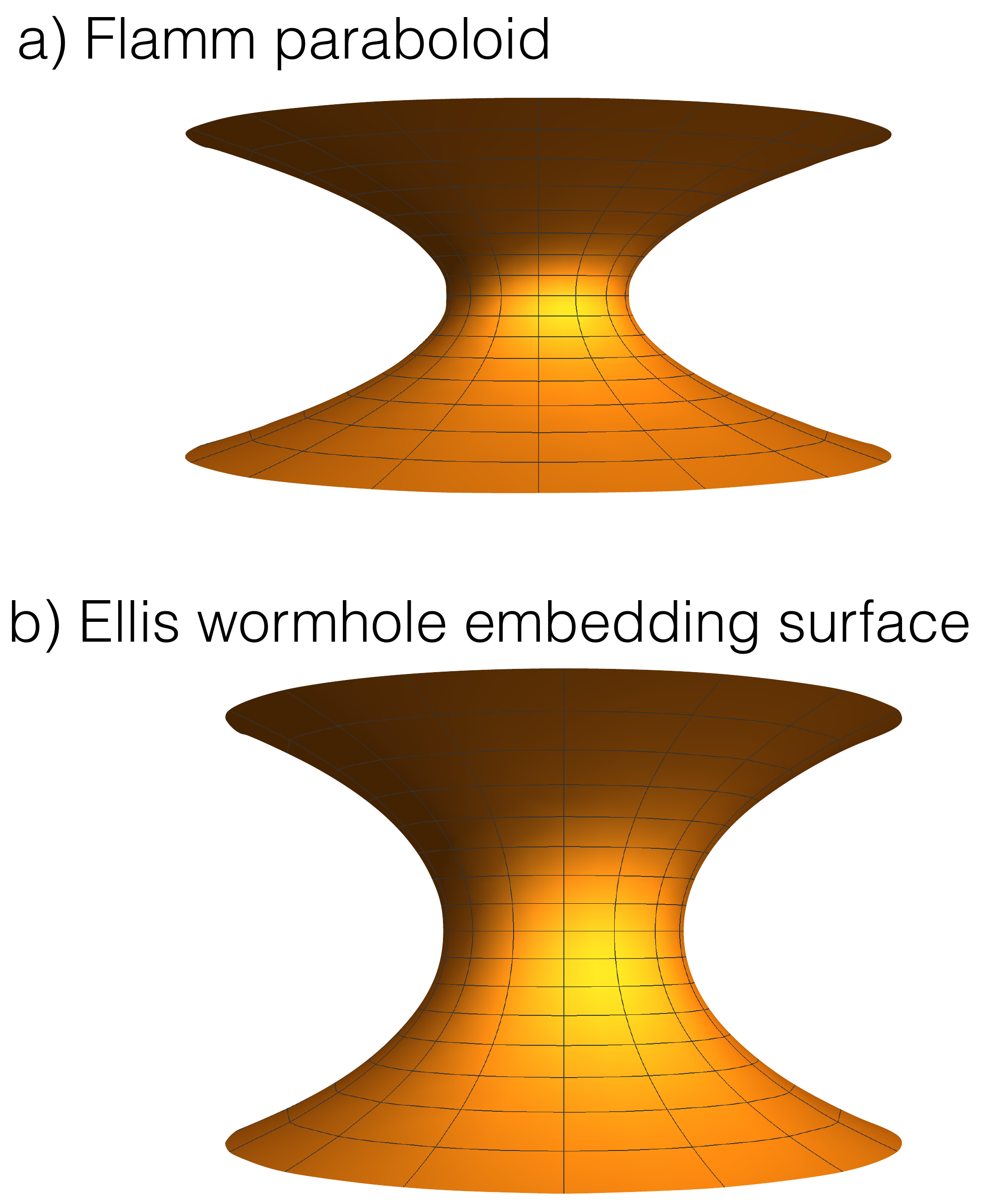}  
\caption{a) Flamm paraboloid corresponding to Schwarzschild black-hole metric. b) Embedding surface of Ellis traversable-wormhole metric. Schwarzschild radius in a) is equal to wormhole's throat radius in b) $b_0=30\, \mu m$. Notice the similarity between both figures. Figure b) does not seem to entail additional experimental challenges. }
   \label{Fig0}
  \end{figure}

In Fig \ref{Fig0}, we compare this surface with Flamm paraboloid, which is the embedding surface of the Schwarzschild black-hole metric and has already been implemented in the lab \cite{rivkanatphot}. We see that the new surface does not seem to entail a significant additional challenge from the experimental viewpoint. 

\section{Wave equations}

Now, we follow the mathematical analysis in \cite{rivkanatphot}. In the coordinate system we are considering, the transverse electric modes do not have $z$ component and, if we write the x component as $E_x= 
\phi(x,z)\xi(h)$, where $h$ is a coordinate normal to the surface, then separation of variables in the corresponding curved-space Maxwell equation for $E_x$ yields \cite{rivkanatphot}: 
\begin{eqnarray}
\frac{1}{\gamma}\partial_x^2\phi+\partial_z^2\phi +\frac{\partial_z\sqrt\gamma}{\sqrt \gamma}\partial_z\phi&=&0\nonumber\\
-\partial_h^2\xi-k^2n^2\xi&=&-q^2\xi,\label{eq:firsteq}
\end{eqnarray}
where $n=c/v$ is the refractive index, $k$ is the wavenumber $k=n\omega/c$ -$\omega$ being the frequency of the wave- and $q$ plays the role of the propagation constant. Finally, using again separation of variables and the k-space representation of the x-part, we write $\phi(x,z)=\frac{1}{\gamma^{1/4}}\psi(z)\int f(k_x)e^{ik_x x}dk_x$ and plug it into \ref{eq:firsteq}, to get an equation which describes propagation along the surface axis \cite{rivkanatphot}:
\begin{equation}
\psi_{zz}=-\frac{1}{16\gamma^2}\left(16q^2\gamma^2+3(\partial_z\gamma)^2-4\gamma\left(4k_x^2+\partial_{zz}\gamma\right)\right)\psi.
\label{eq:evoleqabs}
\end{equation}
Note that $k^2=q^2+k_x^2+k_z^2$.
\section*{Results}

\begin{figure}[h!]\centering
\includegraphics[width=0.9\textwidth]{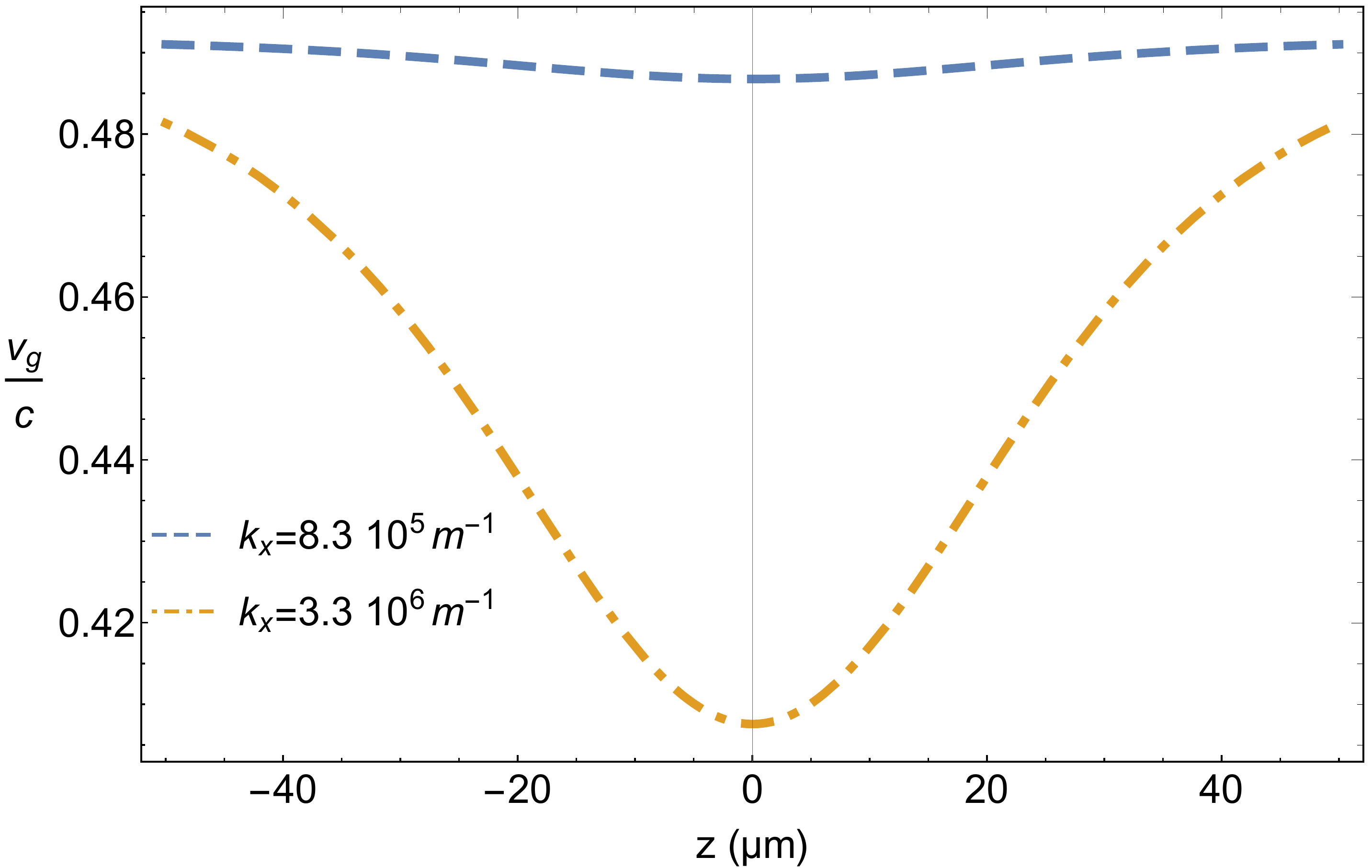}  
\caption{Group velocity $v_g/c$ vs the spatial coordinate $z$ for $b_0= 30\,\mu m$, $k_0=2\pi/\lambda$, $\lambda=780 \operatorname{nm}$, $n_0=1.5$, $q=5.9 \cdot 10^6\operatorname{m^{-1}}$ and the values of $k_x$ indicated. The group velocity can significantly decrease near the wormhole throat at $z=0$.}
   \label{Fig1}
  \end{figure}

\begin{figure}[h!]\centering
\includegraphics[width=0.9\textwidth]{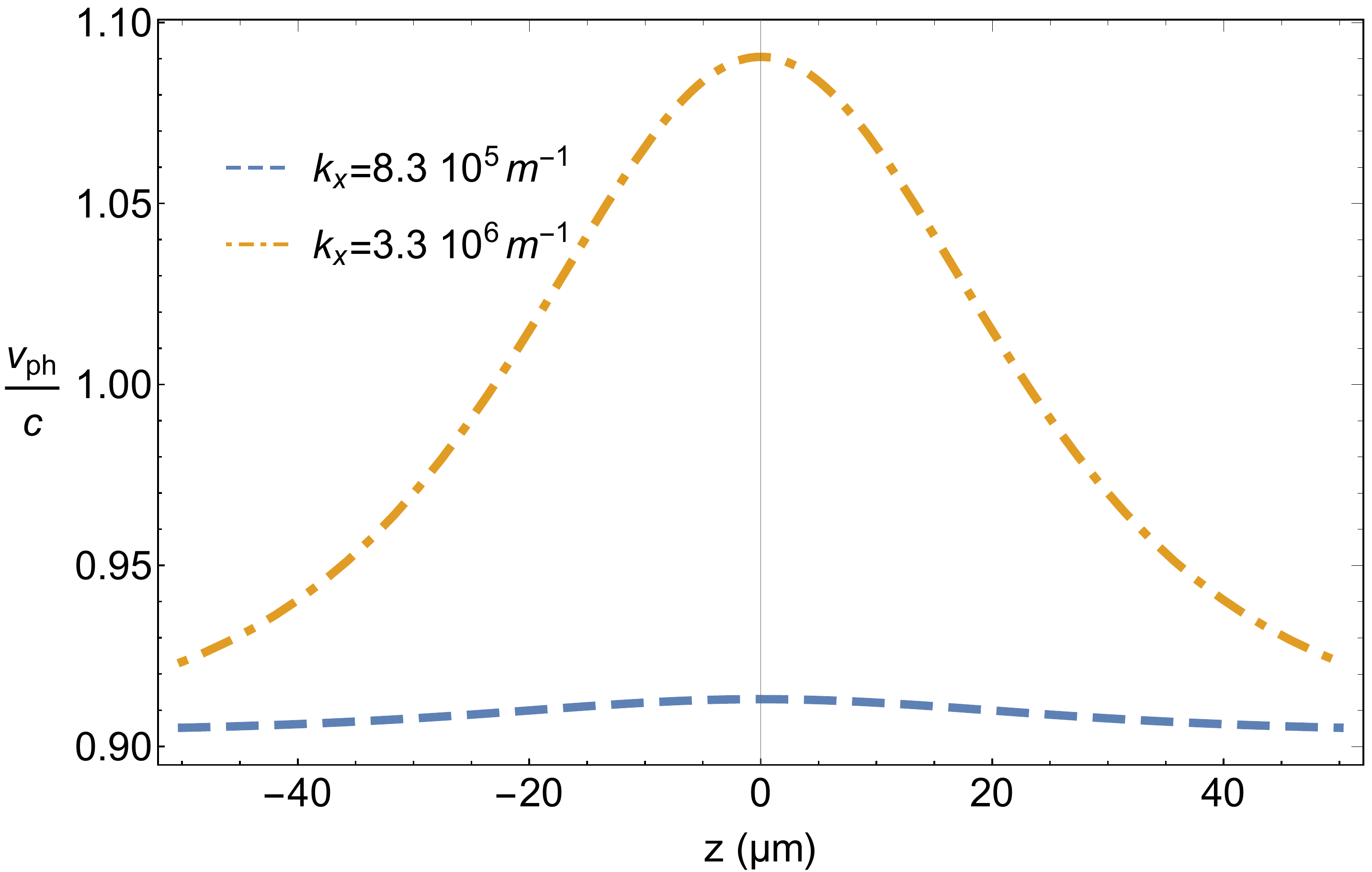}  
\caption{Phase velocity $v_ph/c$ vs the spatial coordinate $z$ for $b_0= 30\,\mu m$, $k_0=2\pi/\lambda$, $\lambda=780 \operatorname{nm}$, $n_0=1.5$, $q=5.9 \cdot 10^6\operatorname{m^{-1}}$ and the values of $k_x$ indicated. The phase velocity increases and can become superluminal near the throat.}
   \label{Fig2}
  \end{figure}

Now plugging Eq. (\ref{eq:gammazdef}) and its derivatives $\partial_z\gamma=2/b_0 \cosh{z/b_0}\sinh{z/b_0}$, $\partial_{zz}=2/b_0^2\large(\sinh^2{(z/b_0\large)}+\cosh^2{\large(z/b_0\large)})$ into Eq. (\ref{eq:evoleqabs}) and making use of hyperbolic trigonometric identities, such as $\cosh^2{x}-\sinh^2{x}=1$ and $1-\tanh^2{x}=\operatorname{sech}^2 x$, we get the equation for $\psi$, describing propagation along the wormhole surface:
\begin{equation}
\psi_{zz}=-\frac{1}{4}\left(4q^2-\frac{1}{b_0^2}-\left(\frac{1}{b_0^2}+4 k_x^2\right)\operatorname{sech}^2\left(\frac{z}{b_0}\right)\right)\psi,
\label{eq:evoleq}
\end{equation}
which, under the WKB approximation has the solution:
\begin{equation}
\psi(z)\simeq\frac{1}{\sqrt[4]{\frac{1}{4}\left(4q^2-\frac{1}{b_0^2}-(\frac{1}{b_0^2}+4 k_x^2)\operatorname{sech}^2\left(\frac{z}{b_0}\right)\right)}}e^{\pm i\,\int\sqrt{\frac{1}{4}\left(4q^2-\frac{1}{b_0^2}-\left(\frac{1}{b_0^2}+4 k_x^2\right)\operatorname{sech}^2\left(\frac{z}{b_0}\right)\right)}dz}.
\label{eq:solwkb}
\end{equation}
Thus, we have:
\begin{equation}
k_z=\sqrt{\frac{1}{4}\left(4q^2-\frac{1}{b_0^2}-\left(\frac{1}{b_0^2}+4 k_x^2\right)\operatorname{sech}^2\left(\frac{z}{b_0}\right)\right)},
\label{eq:kz}
\end{equation}
and we can obtain the phase and group velocities from their definition $v_{ph}=\omega/k_z$, $v_g=\frac{d\omega}{dk_z}$: 
\begin{eqnarray}
v_{ph}&=&\frac{2 c k}{n\sqrt{\left(4q^2-\frac{1}{b_0^2}-(\frac{1}{b_0^2}+4 k_x^2)\operatorname{sech}^2\left(\frac{z}{b_0}\right)\right)}}\nonumber\\
v_{g}&=&\frac{\sqrt{\left(4q^2-\frac{1}{b_0^2}-(\frac{1}{b_0^2}+4 k_x^2)\operatorname{sech}^2\left(\frac{z}{b_0}\right)\right)}}{2 b_0 k n}
\end{eqnarray}
In Figs (\ref{Fig1}), (\ref{Fig2}), we see the behavior of the phase and group velocities
for the same values of $q$ and $k_x$ considered in \cite{rivkanatphot}. We have considered realistic values of $100 \operatorname{\mu m}$ for the length of the surface -centered around the wormhole's throat at $z=0$- and $30 \operatorname{\mu m}$ for its minimum radius -corresponding to the radius of the wormhole throat $b_0$- and a wavelength in the optical regime $\lambda=780 \mu m$.

\begin{figure}[h!]\centering
\includegraphics[width=0.9\textwidth]{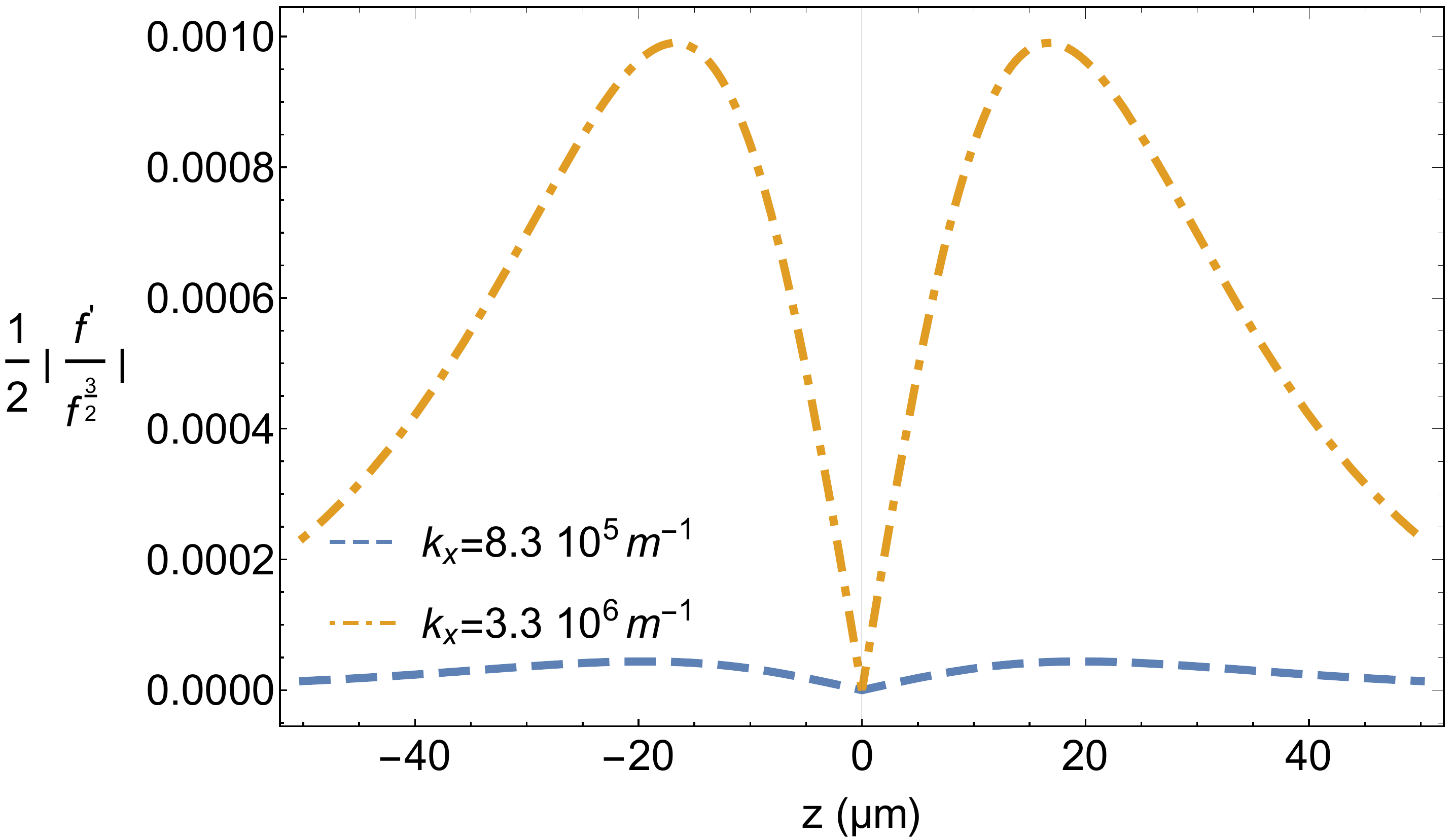}  
\caption{Validity of the WKB approximation $1/2|f'/f^{3/2}|<<1$ vs the spatial coordinate $z$ for $b_0= 30 \,\mu m$, $q=5.9 \cdot10^6\operatorname{m^{-1}}$ and the values of $k_x$ indicated. All the values are the same as in Figs. \ref{Fig1} and \ref{Fig2}.}
   \label{Fig3}
  \end{figure}

In order to check that the WKB approximation is valid, the following condition must be verified:
\begin{equation}
\frac{1}{2}\left | \frac{f'}{f^{\frac{3}{2}}}\right |<<1,
\label{eq:wkbcon}
\end{equation}
where 
\begin{equation}
f= \frac{1}{4}\left(4q^2-\frac{1}{b_0^2}-(\frac{1}{b_0^2}+4 k_x^2)\operatorname{sech}^2\left(\frac{z}{b_0}\right)\right).
\label{eq:f}
\end{equation}
Then, we have:
\begin{equation}
\frac{2}{b_0}\left | \frac{\operatorname{sinh}\left(\frac{z}{b_0}\right)}{\sqrt{\frac{1}{b_0^2}+4 k_x^2}\left (\frac{4q^2-\frac{1}{b_0^2}}{\frac{1}{b_0^2}+4 k_x^2}\operatorname{cosh}^2\left( \frac{z}{b_0}\right)-1\right)^{\frac{3}{2}}}\right | << 1.
\label{eq:wkbcon2}
\end{equation}
In Figur (\ref{Fig3}), we show that the condition in Eq. (\ref{eq:wkbcon2}) is actually well verified for the values of the parameters that we are considering here. As can be seen in Figure (\ref{Fig3}), the approximation is worse as $k_x$ increases. Therefore it would break down for higher values of $k_x$.

\section{Conclusions}

Figs. \ref{Fig1} and \ref{Fig2} display a significant degree of variation of both the group and phase velocities for realistic experimental parameters. The qualitative features are similar to the results in \cite{rivkanatphot}, with the radius of the wormhole throat playing the role of the Schwarzschild radius: group velocity diminishes near the throat, while the phase velocity increases and becomes superluminal. Therefore we show that these features are not a peculiarity of the Schwarzschild geometry, but can be extended to other metrics with different physical interpretation. On the other hand, typical proposals of wormhole simulators are restricted to a one-dimensional section of the spacetime, while here we are considering the full spatial structure. We think that this can be a useful complementary approach to the simulation of wormholes.

In summary, we  have considered light propagation through a 3D nanophotonic structure with the spatial shape of a spacetime containing a traversable wormhole. We show that waves experience significant changes of phase and group velocities when propagating along this particular curved space, in agreement with previous results in Schwarzschild spacetime. We have shown that this proposal of experiment seems to be fully within reach of state-of-the-art nanonphotonics technology. Similar experimental techniques as in \cite{rivkanatphot} would be required, and it does not seem that the embedding surface of the Ellis wormhole entails additional experimental challenges as compared to the Flamm paraboloid of the Schwarzschild metric. Therefore, we conclude that the experimental platform of \cite{rivkanatphot} would be suitable to test our results. However, different nanophotonic platforms \cite{shoresh,nikolaev,malka} with the capability of manipulating the refractive index could be considered. Our results enablee as well the interesting possibility of comparing in the laboratory the propagation of light in wormholes and black holes in order to analyse the potential of propagation features as a means of discriminating between them, which have important consequences in gravitational-wave astronomy \cite{combi,gravastar,konoplya}. Moreover, we foresee that the techniques explained here could be used as well in other spacetimes of interest, for instance spacetimes with exotic properties \cite{sabinexotic}.

\vspace{6pt} 


\acknowledgments{I am indebted to Rivka Bekenstein for useful comments and help. Financial support by Fundaci\'on General CSIC (Programa ComFuturo) and Postdoctoral Junior Leader Fellowship Programme from ``la Caixa'' Banking Foundation is acknowledged.}


\conflictofinterests{The authors declare no conflict of interest.} 




\bibliographystyle{mdpi}

\renewcommand\bibname{References}



\end{document}